\documentclass[epj, nopacs]{svjour}
\usepackage{epsfig}
\usepackage{amssymb}

\newcommand{\be}{\begin{equation}}
\newcommand{\ee}{\end{equation}}
\newcommand{\bea}{\begin{eqnarray}}
\newcommand{\eea}{\end{eqnarray}}
\newcommand{\ov}{\overline}
\newcommand{\ve}{\varepsilon}

\newcommand{\ba}{\begin{array}}
\newcommand{\ea}{\end{array}}

\begin{document}

\title{ V--A sum rules with D=10 operators}
\author{K.N. Zyablyuk}
\institute{Institute of Theoretical and Experimental Physics,
 B.Cheremushkinskaya 25, Moscow 117218, Russia. \email{zyablyuk@itep.ru}}

\abstract{
The difference of vector and axial-vector charged current correlators is analyzed 
by means of QCD sum rules. The contribution of 10-dimensional 4-quark condensates
is calculated and its value is estimated within the framework of factorization hypothesis.
It is compared to the result, obtained from operator fit of Borel sum rules in the complex $q^2$-plane,
calculated from experimental data on hadronic $\tau$-decays. This fit gives accurate values of the light 
quark condensate and quark-gluon mixed condensate. The size of the high-order operators 
and the convergence of operator series are discussed.
}


\maketitle


\section{Introduction}

The QCD sum rules \cite{SVZ} have been widely used for the determination of fundamental 
theoretical parameters, such as the coupling constant $\alpha_s$, quark masses 
and various nonperturbative condensates. Their accuracy depends on experimental
errors and theoretical uncertainties. In many cases both experimental and theoretical
errors are comparable by the order of magnitude, and any improvement is of interest.

In this paper we will consider the 2-point correlators of charged vector and axial-vector
currents, constructed from light $u$,$d$-quarks:
\bea
 && \Pi^U_{\mu\nu}(q)\, = \, i\int dx \, e^{iqx} \left< TU_\mu^\dagger (x) U_\nu (0) \right> \nonumber \\
 &&  \, = \, (q_\mu q_\nu - g_{\mu\nu} q^2 )\,\Pi^{(1)}_U (q^2) \,+\,q_\mu q_\nu \, \Pi^{(0)}_U(q^2)
\label{2ccdef}
\eea
where 
$$
 U=V,A: \qquad V_\mu={\bar u}\gamma_\mu d  \; , \;
 A_\mu ={\bar u}\gamma_\mu\gamma^5 d \; .
$$
The polarization functions $\Pi^{(i)}(s)$ have a cut along the real axes in the complex 
$s=q^2$ plane. Their imaginary parts (spectral functions)
\bea
v_1/a_1(s) & = & 2\pi \,{\rm Im} \,\Pi^{(1)}_{V/A} (s+i0)   \; , \nonumber \\
a_0(s) & = & 2\pi \,{\rm Im} \,\Pi^{(0)}_A (s+i0) 
\eea
have been measured for $0<s<m_\tau^2$ by ALEPH \cite{ALEPH} and OPAL \cite{OPAL} collaborations
from hadronic decays of $\tau$-lepton.

\begin{figure}[tb]
 \epsfig{file=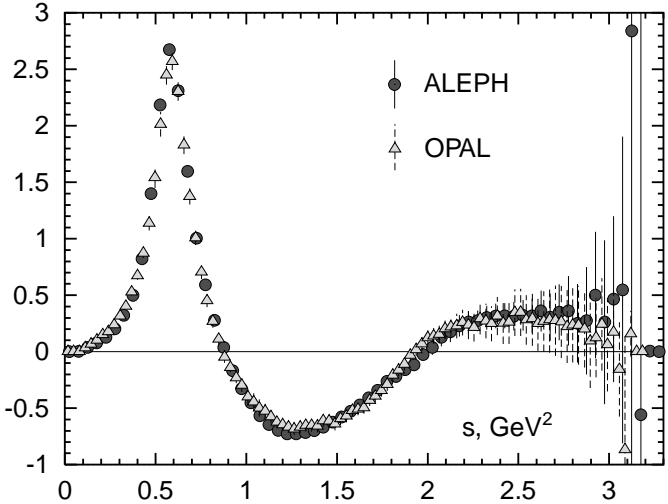, width=88mm}
 \caption{Spectral function $v_1(s)-a_1(s)$ obtained from ALEPH \cite{ALEPH} and 
 OPAL \cite{OPAL} data}
 \label{vma_exp}
 \end{figure}

Of particular interest  is the difference $\Pi_V^{(1)}-\Pi_A^{(1)}$, since
it does not contain any perturbative contribution in the massless quark limit.
The experimental data on the difference $v_1(s)-a_1(s)$ are shown in Fig.~\ref{vma_exp}. 
As demonstrated in \cite{IZ}, the dispersion relation can be written in the following form:
\bea
\Pi_V^{(1)}(s)-\Pi_A^{(1)}(s) & = & {1\over 2\pi^2}\int_0^\infty {v_1(t)-a_1(t) \over t-s} \,dt \,+ \,
{f_\pi^2 \over s} \nonumber \\
 & = &\sum_{D\ge 4} {O_D^{V-A}\over (-s)^{D/2}}
\label{disr}
\eea
where the sum goes over even dimensions $D$ of the operators (condensates) 
$O_D$. The term $f_\pi^2\over s$, $f_\pi=130.7\,{\rm MeV}$ is the pion decay constant,
 is the kinematical pole of the axial polarization function $\Pi^A_{\mu\nu}$, see \cite{IZ} for details.
In (\ref{disr}) and below the notation $O_D^{V-A}$ stands for the condensates 
with all $\alpha_s$ corrections, including slowly varying logarithmic terms $\sim \ln^n(-s)$.
The list of the condensate contribution to the vector and axial correlators separately can be 
found in \cite{BNP}.

The sum rules for the difference (\ref{disr}) have been studied in \cite{IZ}, \cite{DGHS}--\cite{RL}
where the lowest order condensates $O_D^{V-A}$ were found.  Although the published values
of $O_6^{V-A}$ are close to each other (within the errors), this is not the case for
the operator $O_8^{V-A}$. In \cite{IZ}, \cite{DGHS} positive values of the $D=8$ condensate
were found, but the authors of recent publications \cite{CGM}, \cite{RL} have obtained negative
condensate $O_8^{V-A}$. The source of this discrepancy could be very large 
condensates of dimension $D=10$ and higher, accounted  in \cite{CGM}, \cite{RL}: a typical 
ratio of the condensates in these papers is $|O_{2n+2}/O_{2n}|\sim 5-10 \,{\rm GeV}^2$. 
If this statement is correct, the OPE analysis of \cite{IZ} would be invalid, because the
contribution of unknown high-order terms was estimated from the assumption
$|O_{10}/O_6| \lesssim 1-2\,{\rm GeV}^4$. For this reason it would be interesting to 
find the operator $O_{10}^{V-A}$ independently and compare it with the 
sum rule results. 

In this paper we repeat the analysis of \cite{IZ} with the $D=10$ operator included.
In Section 2 all necessary $V-A$ operators, obtained from the Operator Product Expansion in QCD,
are listed and their values are estimated within the framework of the factorization 
hypothesis. In Section 3 the operator values are obtained from the fit to Borel sum rules.
In the last Section the validity of our assumptions is discussed and the results are compared
with the ones obtained in other publications. The complete form of the $D=10$ operator
and technical details of its derivation are dropped to Appendices A,B.

\section{V--A operator expansion}

The first term in the operator series (\ref{disr}) is the $D=4$ operator:
\bea
 O_4^{V-A} & = & 2(m_u+m_d)\langle {\bar q}q\rangle \nonumber \\
 & \times &  \left[\, 1\,+ \,{4\over 3}\, {\alpha_s(Q^2)\over \pi} +  \,{59\over 6} 
 \left( {\alpha_s(Q^2)\over \pi} \right)^2\, \right]
 \label{o4vma}
\eea
where $Q^2=-q^2$, we assume $\langle {\bar u}u\rangle =\langle {\bar d}d\rangle \equiv
 \langle {\bar q}q\rangle $. The $\alpha_s$ corrections have been computed in \cite{Gen} and \cite{CGS}.
In fact, the contribution of the $D=4$ operator to the sum rules considered here, is small. 
So we can safely neglect the $\alpha_s$-corrections in (\ref{o4vma}) and put 
$O_4^{V-A}=-f_\pi^2 m_\pi^2=-3.3\times 10^{-4} \, {\rm GeV}^4$, as follows from Gell-Mann-Oakes-Renner
low energy theorem \cite{GMOR}.

The $D=6$ operator in factorized form is equal to:
\be
O_6^{V-A} = -8\pi C_N \alpha_s \langle {\bar q}q\rangle^2 \left[ 1+{\alpha_s (\mu^2)\over\pi}
\left( c_6 -{1\over 4}\ln{Q^2\over\mu^2} \right) \right]
\label{o6vma}
\ee
where $C_N=1-N_c^{-2}=8/9$ is the color factor, which appears in the factorization 
of the 4-quark operators at the leading $\alpha_s$-order. The NLO terms were computed
in \cite{LSC} and the constant $c_6$ was found equal to $247/48$. In \cite{AC} 
another treatment of $\gamma^5$ matrix in dimensional regularization was employed, leading 
to $c_6=89/48$. For the later choice at $\mu=1\,{\rm GeV}$ and $\alpha_s(\mu^2)=0.5$
one finds the factor in square brackets in (\ref{o6vma}) equal to 1.3 (the logarithmic 
term can be neglected due to small numerical coefficient).

The contribution of the $D=8$ 4-quark condensates to the vector current correlator
was originally obtained in \cite{DS} in factorized form and in \cite{GP} in complete (nonfactorized)
form. In \cite{IZ} these results were verified and an ambiguity of the factorization 
at the $N_c^{-2}$ order was pointed out. Here we will follow the factorization procedure,
described in Appendix B. The result is\footnote{In \cite{IZ} the factor $C_N$ was ignored,
since $O(N_c^{-2})$ terms were neglected}:
\be
O_8^{V-A}\,=\, 8\pi C_N \alpha_s m_0^2 \langle {\bar q}q\rangle^2 \; ,
\label{o8vma}
\ee
where the mass $m_0$ is defined from the 5-dimensional quark-gluon mixed condensate:
\be
 i\langle {\bar q} \hat{G} q\rangle\,= 
2\langle {\bar q} D^2 q \rangle \,=\, -\, m_0^2 \langle {\bar q}q\rangle
\ee
where $\hat{G}=\gamma_\alpha \gamma_\beta G_{\alpha\beta}$,  $G_{\alpha\beta}=i[D_\alpha, D_\beta]$
is the gluon field strength, see Appendix A for more definitions. 
The parameter $m_0^2$ has a meaning of typical momentum of virtual quarks in vacuum.
It was found from baryonic sum rules  $m_0^2=0.8\pm 0.2 \, {\rm GeV}^2$ \cite{BI,DJN},
and $B-B^*$ splitting \cite{Nar}. The values close to $1\,{\rm GeV}^2$ were also obtained  
from the latest lattice calculation \cite{ChH} and in QCD string model \cite{DiGS}.

\begin{figure}[tb]
  \epsfig{file=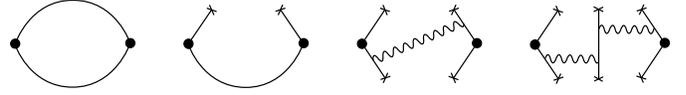, width=88mm}
 \caption{Condensate expansion (\ref{c10exp}) by the number of quarks in vacuum. Circles stand for 
the currents $V/A$, crosses are quarks in vacuum; gluons in vacuum are not shown.}
 \label{diag_10}
 \end{figure}

There are many different condensates of dimension $D=10$. They can be grouped into 
four parts:
\be
O_{10}^{V-A}\,=\,O^{(0)}_{10}\,+\,O^{(2)}_{10}\,+\,O^{(4)}_{10}\,+\,O^{(6)}_{10}
\label{c10exp}
\ee
where upper index $(i)$ denotes the number of quarks in vacuum. This separation 
is shown diagrammatically in Fig.~\ref{diag_10}. The purely gluonic operators $O^{(0)}$ and
the 2-quark ones $O^{(2)}$ cancel in the $V-A$ correlator in the limit of massless $u,d$-quarks.
The operators with 6 quarks in vacuum have the structure 
$\langle ({\bar q}q)^2 ({\bar q} D q) \rangle$. After factorization they become 
$\sim m\langle {\bar q}q\rangle^3 $, which is again negligible for light quarks.
The only essential contribution to the $V-A$ sum rules
comes from the 4-quark operators $O_{10}^{(4)}$. 

In this paper we have computed the contribution of the 4-quark condensates to the vector
and axial current correlator. Details of calculation and complete form of the operator
$O_{10}^{(4)}$ are given in Appendix A. The factorization scheme necessary to reduce large
number of independent structures, is described in Appendix B. The result is:
\bea
O_{10}^{V-A} & = & \pi\alpha_s C_N \left[ \,{50\over 9}\,\langle {\bar q} \hat{G} q\rangle^2 \right. \nonumber \\
 & & \left. - \,16\left( \,3\,X_1\,-\,X_2\,+\,X_3\,+\,{7\over 6}\,X_4\, \right)\langle {\bar q}q\rangle \, \right]
\label{o10vma}
\eea
where $X_i$ are 4 independent $D=7$ quark-gluon condensates:
\bea
 && X_1  = \langle {\bar q} G_{\alpha\beta}G_{\alpha\beta} q\rangle  , \qquad \; \;
 X_2  =i\langle {\bar q}\gamma^5 G_{\alpha\beta}{\tilde G}_{\alpha\beta} q\rangle  , \; \nonumber \\
 && X_3  =\langle {\bar q} \gamma_{\alpha\beta}G_{\alpha\gamma}G_{\beta\gamma} q\rangle, \quad
 X_4  =i\langle {\bar q}\gamma_{\alpha\beta}(D_\alpha J_\beta)  q \rangle  .
\label{d7cond}
\eea
where $\tilde{G}_{\alpha\beta}$ is the dual gluon-field strength, 
$\gamma_{\alpha\beta}={1\over 2}(\gamma_\alpha \gamma_\beta -\gamma_\beta \gamma_\alpha )$ 
and $J_\alpha = D_\beta G_{\alpha\beta}$.
Their numerical values are not known. The condensate $X_4$ can be brought to the 4-quark form
$X_4\sim \langle ({\bar q}q) ({\bar q} D q) \rangle \sim m\langle {\bar q}q\rangle^2 $, which is negligible.
In order to estimate other condensates, we assume further factorization according to 
$\langle {\bar q} \Gamma q \rangle = \langle {\bar q} q\rangle \, \langle {\rm tr} \, \Gamma \rangle /(4N_c)$,
the trace is taken both over color and spinor indices. Then:
\be
X_1\,=\,{2\pi\over 3}\, \alpha_s \left< G_{\alpha\beta}^a G_{\alpha\beta}^a\right>
  \langle {\bar q} q\rangle  \; , \qquad X_2\,=\,X_3\,=\,X_4\,=\,0 \; .
\ee
Under these assumptions the operator (\ref{o10vma}) takes the form:
\be
O_{10}^{V-A} \,=\, -\,\pi\alpha_s C_N \langle {\bar q} q\rangle^2 \left[ \,{50\over 9}\, m_0^4 \,+\,
 32 \pi \alpha_s  \left< G^2 \right> \, \right]
\label{o10vma2}
\ee
It is rather difficult to find accurate value of the gluon condensate
from any sum rule. Detailed analysis of charmonium
sum rules performed in \cite{IZ2} has lead to the restriction
$\langle {\alpha_s\over \pi} G^2\rangle = 0.009\pm 0.007 \,{\rm GeV}^4$, in agreement with 
many previous estimations. Taking this central value and $m_0^2=1\,{\rm GeV}^2$,
one obtains $O_{10}^{V-A}/O_6^{V-A}=0.8\,{\rm GeV}^4$. For 
$O_6^{V-A}=-(6.8\pm 2.1)\times 10^{-3}\,{\rm GeV}^6$
\cite{IZ} we find the following estimation of the $D=10$ $V-A$ condensate:
\be
O_{10}^{V-A} \,=\,-5\times 10^{-3}\,{\rm GeV}^{10}
\label{o10num}
\ee
In the next section we will compare this estimation with results of the fit, obtained from the sum rules.


\section{V--A sum rules}

Many different sum rules have been investigated in order to determine numerical values of the
condensates. Most of authors employ polynomial sum rules: the correlator
$\Pi^{(1)}_V(s)-\Pi^{(1)}_A(s)$ is multiplied on some polynomial of $s$ and then integrated
over the circle $|s|=s_0$ in the complex $s$-plane. Their advantages are: 1) 
one does not need to know the spectral function $v_1(s)-a_1(s)$ for $s>s_0$, which 
allows to reduce high error from the region $s\approx m_\tau^2$ by choosing $s_0$ reasonably
below $m_\tau^2$ and 2) all operators of dimension higher then the polynomial dimension,
do not enter these sum rules due to Cauchy theorem. But the disadvantages are also obvious. If the 
operator expansion (\ref{disr}) is divergent (asymptotic), the Cauchy theorem is not applicable
to this series. Moreover, possible logarithmical terms $\sim \ln^k\!Q^2/Q^{2n}$ appear at the
NLO in $\alpha_s$ expansion. These terms contribute to any polynomial sum rules.
It makes uncontrollable the contribution of the high order operators to the 
polynomial sum rules at $s_0\lesssim 2 \, {\rm GeV}^2$, especially for large
ones as obtained in \cite{CGM,RL}.

For these reasons  we prefer Borel sum rules, where the high order operators are suppressed
as $O_{2n}/n!$. In order to separate out the contributions of different 
operators from each other, one may consider the Borel transformation in the complex plane of the 
Borel mass $M^2\to M^2e^{i(\pi-\phi)}$ (which is equivalent to the Borel operator
applied to the dispersion relation (\ref{disr}) written  along the ray $s\to se^{i\phi}$ in the
complex $s$-plane \cite{IZ}). The real and imaginary parts of the Borel transformation are:
\bea
 && \int_0^{s_m}\!\exp{\left( {s\over M^2}\cos{\phi}\right)} \cos{\left({s\over M^2}\sin{\phi}\right)}
\, (v_1-a_1)(s)\, {ds\over 2\pi^2}  \nonumber \\
 && \qquad =\,f_\pi^2 +\sum_{k=1}^\infty (-)^k 
 {\cos{(k\phi)} \, O_{2k+2}^{V-A} \over k!\, M^{2k} } \label{borre}  \\
 && \int_0^{s_m}\!\exp{\left( {s\over M^2}\cos{\phi}\right)} \sin{\left({s\over M^2}\sin{\phi}\right)}
 \, (v_1-a_1)(s)\, {ds\over 2\pi^2 M^2} \nonumber \\
 && \qquad =\,\sum_{k=1}^\infty (-)^k 
 {\sin{(k\phi)} \, O_{2k+2}^{V-A} \over k! \, M^{2k+2}} \label{borim}
\eea
We made the imaginary part (\ref{borim}) dimensionless, while the real part (\ref{borre})
has dimension ${\rm GeV}^2$ in order to separate out the leading constant term $f_\pi^2$.
The logarithmical terms are neglected in the rhs of (\ref{borre},\ref{borim}), 
otherwise the terms $\sim\ln{M^2}$ appear. The only known logarithmical  term is 
in the $\alpha_s$-correction to the $D=6$ operator
(\ref{o6vma}). It can be easily taken into account (see \cite{CGM} for explicit formulae), 
but its relative contribution is negligible due to small numerical factor, so we shall ignore it.

The derivation of (\ref{borre},\ref{borim}) from the dispersion relation (\ref{disr})
implies infinite upper integration limit $s_m=\infty$. Experimental data on the axial function $a_1(s)$ 
are available only for $s<m_\tau^2=3.16\,{\rm GeV}^2$. However the data at $s>3\,{\rm GeV}^2$ are
rather unstable and have large error because of low statistics, see Fig \ref{vma_exp}. 
For this reason we put $s_m=3.0 \, {\rm GeV}^2$ in (\ref{borre},\ref{borim}). Removal of the data above 
this point does not change the Borel transform 
significantly (if $M^2$ is not sufficiently large), but may reduce the errors. 
In fact, the sum rules considered here do not rely on the high-energy data: say,
if the upper integration limit $s_m$ is reduced to $2.5\, {\rm GeV}^2$, the condesates change 
at most within  $10\%$ limit. If the data above $3\,{\rm GeV}^2$ are removed, both ALEPH and OPAL
data give almost equal central values and similar errors of the Borel transforms (\ref{borre},\ref{borim}). 
For this reason we will psesent below the analisys of ALEPH data only, since they have smaller errors. 
The condensates,  obtained from OPAL data are almost the same.

The argument of the exponent must be negative $\cos{\phi}<0$ 
in order to suppress contribution of the high-energy states from unknown region $s>m_\tau^2$,
which means $\pi/2<\phi<\pi$. Of special interest are the closest to $\pi$ (minimal error)
angles at which the contribution of some operator $O_{2k+2}$ vanishes. Such angles are
$\phi=\pi(2k-1)/(2k)$, $k=2,3,\ldots$ for real part (\ref{borre}) and 
$\phi=\pi(k-1)/k$, $k=3,4,\ldots$ for imaginary one (\ref{borim}). 
The sum rules (\ref{borre},\ref{borim}) at some of these angles were considered in \cite{IZ}
with the operators $O_6$ and $O_8$ as free parameters to fit. It was shown, that 
for $O_6^{V-A}=-(6.8\pm 2.1)\times 10^{-3}\,{\rm GeV}^6$ and
$O_8^{V-A}=(7\pm 4)\times 10^{-3}\,{\rm GeV}^6$ they are well satisfied for
$M^2 > 0.6\,{\rm GeV}^2$. 

\begin{table*}
$$
 \ba{c|c|ccc|c}\hline
\phi &\; M^2,\,{\rm GeV}^2 \; & O_6^{V-A} & O_8^{V-A} & O_{10}^{V-A} & \chi_0^2  \\ \hline\hline
2\pi/3 \; & \ba{c} 0.4-1.0 \\ 0.5 -1.0 \ea &
    \; \ba{c} -7.0\pm 1.4\\ -7.0\pm 2.5 \ea \; 
      & &\; \ba{c} -3.8\pm 3.3 \\ -3.9\pm 10.3 \ea\; 
               & \; \ba{c} 0.14 \\ 0.14 \ea  \\ \hline
3\pi/4 \; & \ba{c} 0.4-1.0\\  0.5-1.0 \ea 
    &\; \ba{c} -7.3\pm 1.1\\ -7.9\pm 2.3 \ea \; 
            & \; \ba{c} 8.0\pm 2.1\\ 9.5\pm 2.5\ea\; &   & \; \ba{c} 0.17 \\ 0.12 \ea \\ \hline
4\pi/5 \; & \ba{c} 0.3-1.0 \\ 0.4-1.0 \ea  
    & \; \ba{c} -8.1\pm 1.8\\ -9.3\pm 5.1 \ea \; 
       & \; \ba{c} 10.3\pm 4.1\\ 13.8\pm 15.1\ea \; 
          &\; \ba{c} -7.6\pm 5.0\\ -13.2\pm 23.8\ea \; & \; \ba{c} 0.11 \\ 0.06 \ea  \\ \hline \hline
 {\rm all} & &\; \ba{c} -7.2\pm 1.2\\ -7.5\pm 2.3 \ea \; & \; \ba{c} 7.8\pm 2.5\\ 8.6\pm 6.0\ea \;&
 \; \ba{c} -4.4\pm 2.8\\ -5.2\pm 8.4\ea \;  & \; \ba{c} 0.40 \\ 0.20 \ea \\ \hline
\ea
$$
\caption{Operator fit obtained from eq (\ref{borim}) for different angles $\phi$;
operators are in $10^{-3}\,{\rm GeV}^D$. 
Last two lines contain combined fits for all these angles for upper/lower choice of $M^2$ range.}
\label{bi_tabl}
\end{table*}

It is more difficult to find high-order condensates (say $O_{10}^{V-A}$) from the sum rules,
since several unknown parameters enter the same equation and the high-order condensate 
strongly depends on exact values of the low-order ones. Here one needs to consider 
the Borel transformation at several values of $M^2$, where the relative contributions 
of various condensates are different. In other words, we may fit the shape of theoretical curve
with experimental one within some reasonable region $M_1^2<M^2<M_2^2$. 
For this purpose it is natural to define the least square deviation, normalized to experimental error:
\be
\label{chi2}
\chi^2\,=\,{1\over M_2^2-M_1^2} \int_{M_1^2}^{M_2^2} dM^2 \, 
\left( {B^{theor} - B^{exp}\over \Delta B^{exp}} \right)^2
\ee
where $B^{theor}/B^{exp}$ is the right/left hand side of the Borel sum rules (\ref{borre},\ref{borim}).
One may calculate $\chi^2$ with theoretical condensates $O_i$ as free parameters. It is
quadratic function of them:
\be
 \chi^2\,=\,\chi^2_0 \, + \,\sum_{i,j} C_{ij}\left( O_i-\ov{O}_i \right) \left( O_j-\ov{O}_j \right)
\ee
Obviously $\ov{O}_i$ are the central values of the condensates.  According to the definition (\ref{chi2})
it is natural to consider equation $\chi^2=1$ as the one, which determines the border of the 
$1\sigma$ deviation area in the parameter space. Diagonalizing the matrix $C_{ij}$ by means of 
orthogonal rotation we conclude, that $C_{ij}$ is inverse to the covariance matrix
$\ov{\Delta O_i \cdot \Delta O_j}=(C^{-1})_{ij}$. For a good fit $\chi_0^2\ll 1$.

The fit results depend on the Borel mass limits $M^2_{1,2}$ in (\ref{chi2}). 
For $M^2>1\,{\rm GeV}^2$ the experimental errors are large, so we take $M^2_2=1\,{\rm GeV}^2$. 
The lower limit $M_1^2$ depends on the size of neglected high order operators. 
In \cite{IZ} a good coincidence of experimental
and theoretical curves was observed for $M^2>0.6\,{\rm GeV}^2$. Here we include the operator
$O_{10}$ in the analysis, so this value can be slightly reduced. As follows from our calculation
of the 4-quark condensates (\ref{o6vma}), (\ref{o8vma}), (\ref{o10vma2}), it is reasonable
to assume $O_{2n+2}/O_{2n}\sim m_0^2 \approx 0.7 \,{\rm GeV}^2$. It leads to an estimation
$|O_{12}^{V-A}| \sim 3 \times 10^{-3} \, {\rm GeV}^{-12}$ which allows us to take 
$M^2_1=0.4\,{\rm GeV}^2$, where typical contribution of such operator is not higher than $20\%$.
At the angles, where the contribution of the operator $O_{12}$ 
vanishes, the Borel mass can be reduced even further, say, to $M_1^2=0.3\,{\rm GeV}^2$. 
All these assumptions are confirmed by the results of the fit, see Figures below. 

\begin{figure}[tb]
  \epsfig{file=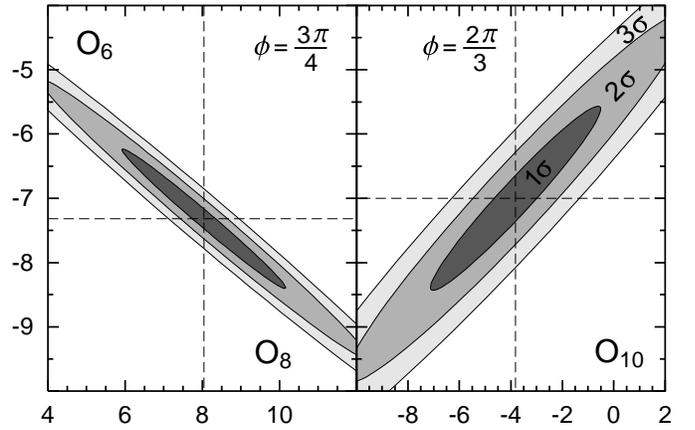, width=88mm}
 \caption{Confidence level contours, obtained from 2-parameter fits of the sum rule (\ref{borim})
in the range $M^2=0.4-1\,{\rm GeV}^2$. The condensates are in $10^{-3}\,{\rm GeV}^D$, the contours show
$1,2,3\sigma$ deviations of $\chi^2$.}
 \label{fig_osig}
 \end{figure}

The condensates, obtained from real part of Borel transformation (\ref{borre}), are
sensitive to exact value of $f_\pi$. For this reason we shall use the imaginary part (\ref{borim})
for numerical fit. The best angles are $\phi=2\pi/3, 3\pi/4, 4\pi/5$ where the contribution
of the operators $O_8, O_{10}, O_{12}$ vanishes respectively. The fit results for each angle are
summarized in the Table \ref{bi_tabl}.  The lowest errors are obtained from the 2-parameter fits
at the first two angles. The deviation $\chi_0^2$ for these fits is sufficiently small. For this reason 
the inclusion of additional parameters, say $O_{12}$, will not improve the fit quality, but will
increase the errors only.

The operator values, obtained from the sum rules, are not independent but have large covariances 
$$
\rho_{ij}=\ov{\Delta O_i \, \Delta O_j}/(\ov{(\Delta O_i)^2}\,\,\ov{(\Delta O_j)^2})^{1/2}. 
$$
All fits give $\rho_{6,10}\approx 1$ and $\rho_{6,8}\approx\rho_{8,10}\approx -1$. 
For the 2-parameter fits the covariances can be demonstrated on the confidence level plots, see 
Fig \ref{fig_osig}. The equations $\chi^2=n^2$ set the ellipses, which are the borders of the $n\sigma$
deviation area.

One may also try to fit the condensates at all these angles simultaneously by minimizing 
$\chi^2_{all}={1\over 3}[\chi^2(2\pi/3)+\chi^2(3\pi/4)+\chi^2(4\pi/5)]$,
see the last two lines in the Table. As the final result of our analysis we take this combined fit: 
\bea
O_6^{V-A} & = & -\,(7.2\pm 1.2)\times 10^{-3} \,{\rm GeV}^6  \nonumber \\
O_8^{V-A} & = & \; \; \; \, (7.8\pm 2.5)\times 10^{-3} \,{\rm GeV}^8 \nonumber \\
O_{10}^{V-A} & = & -\, (4.4\pm 2.8)\times 10^{-3} \,{\rm GeV}^{10} 
\label{combfit}
\eea 
The lower limit of the Borel mass in (\ref{chi2}) was taken $M_1^2=0.4\,{\rm GeV}^2$ for the first two 
angles and $M_1^2=0.3\,{\rm GeV}^2$ for the last one. If $M_1^2$ is taken by $0.1\,{\rm GeV}^2$ higher,
the errors are increased, especially for the high dimension operators, see the last line in the Table.

\begin{figure}[tb]
\hspace{8mm}\epsfig{file=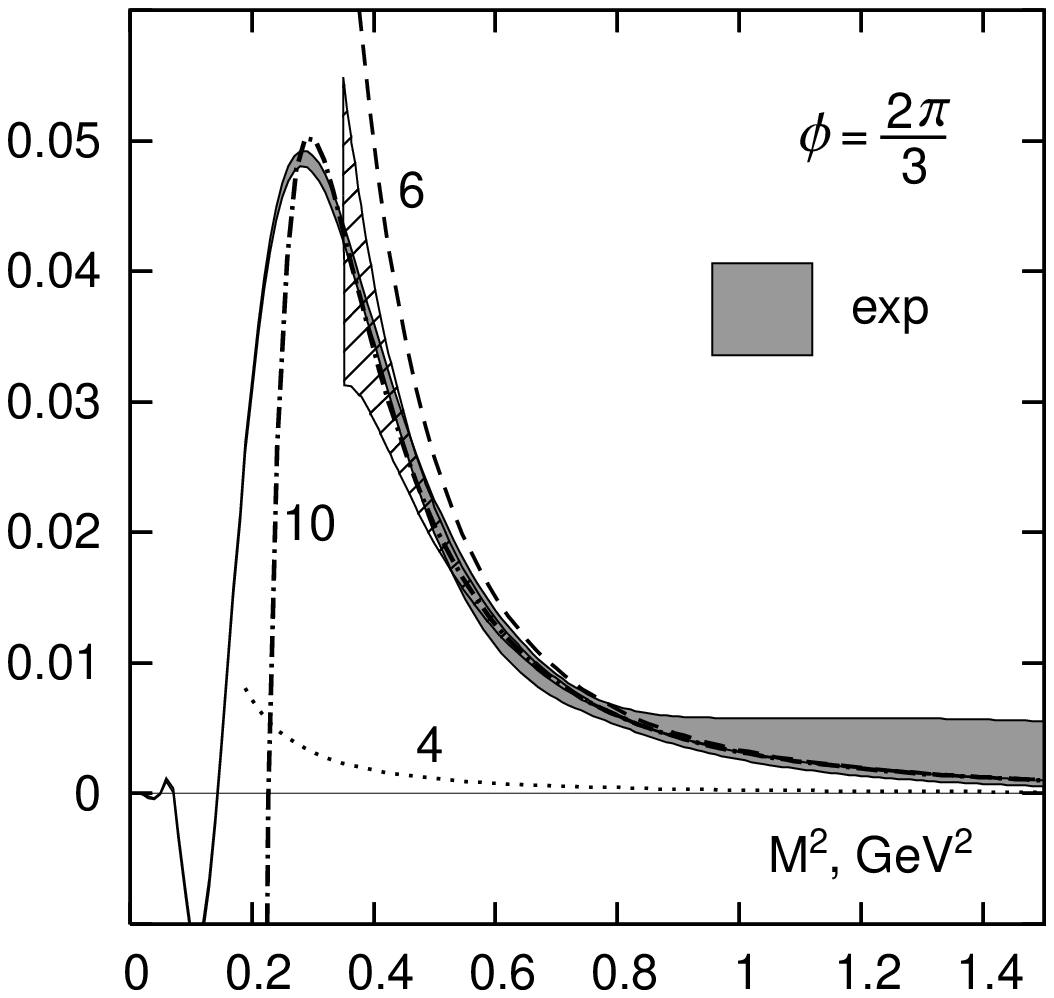, width=70mm} 

\vspace{3mm}\hspace{8mm}\epsfig{file=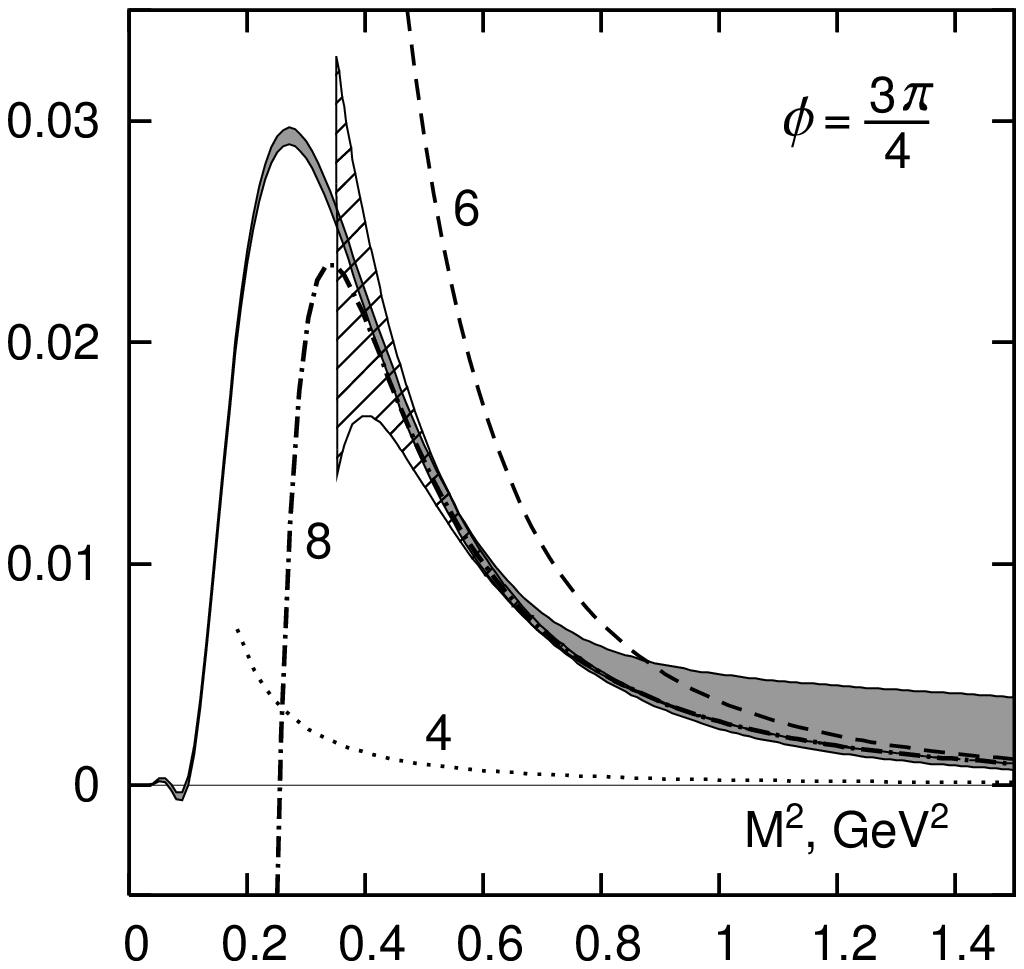, width=70mm}
 \caption{Imaginary part of the Borel transformation  for $\phi=2\pi/3$ (no $O_8$)
and $\phi=3\pi/4$ (no $O_{10}$). Shaded area is the l.h.s.~of (\ref{borim}) calculated from 
experimental data (with error). The lines display the operator series in the r.h.s.~of (\ref{borim})
with condensates, equal to the central values of (\ref{combfit}).
The number nearby each line shows the order of the series; say "8" denotes the contribution 
$O_4+O_6+O_8$. Grid shows possible contribution 
of the operator $O_{12}$ within the limits $|O_{12}^{V-A}|<3\times 10^{-3}\,{\rm GeV}^{12}$.}
 \label{bs_12}
 \end{figure}

\begin{figure}[tb]
\hspace{8mm}\epsfig{file=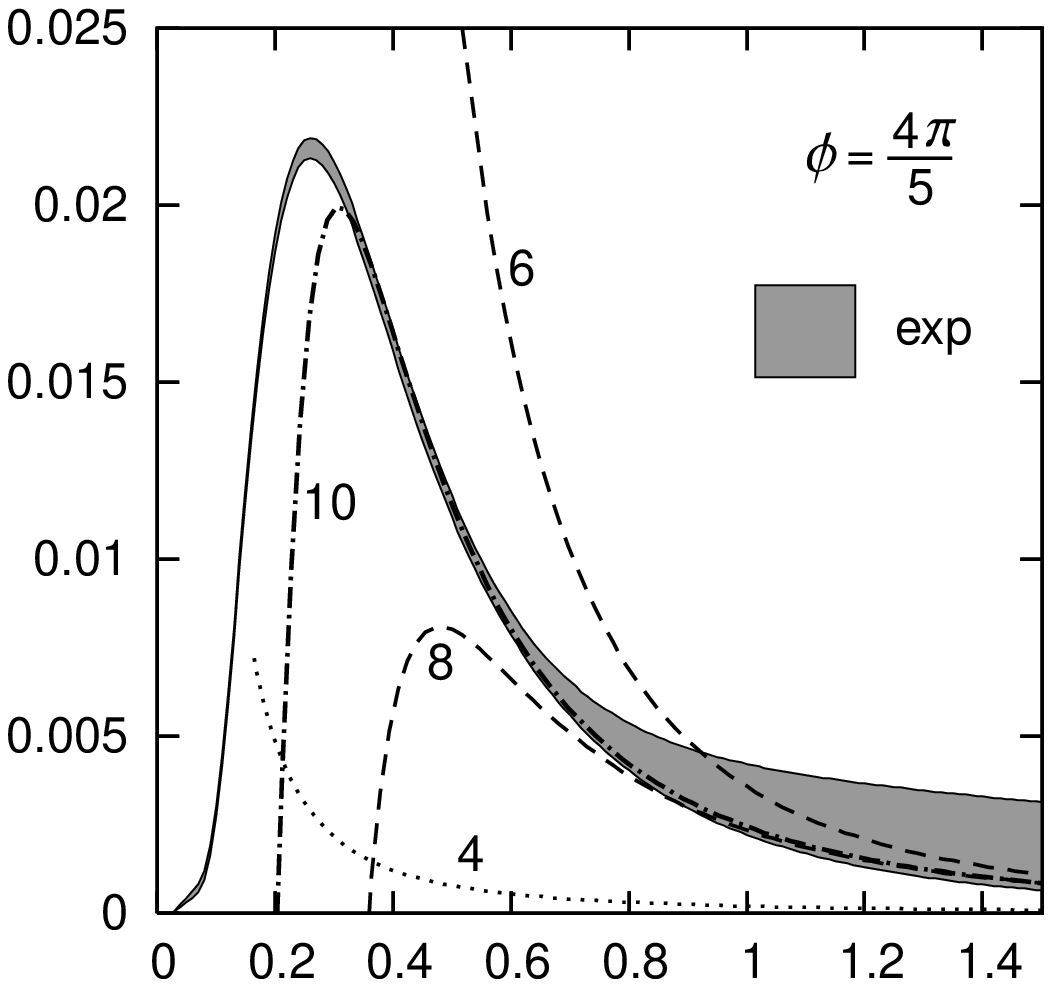, width=70mm}

\vspace{3mm}\hspace{8mm}\epsfig{file=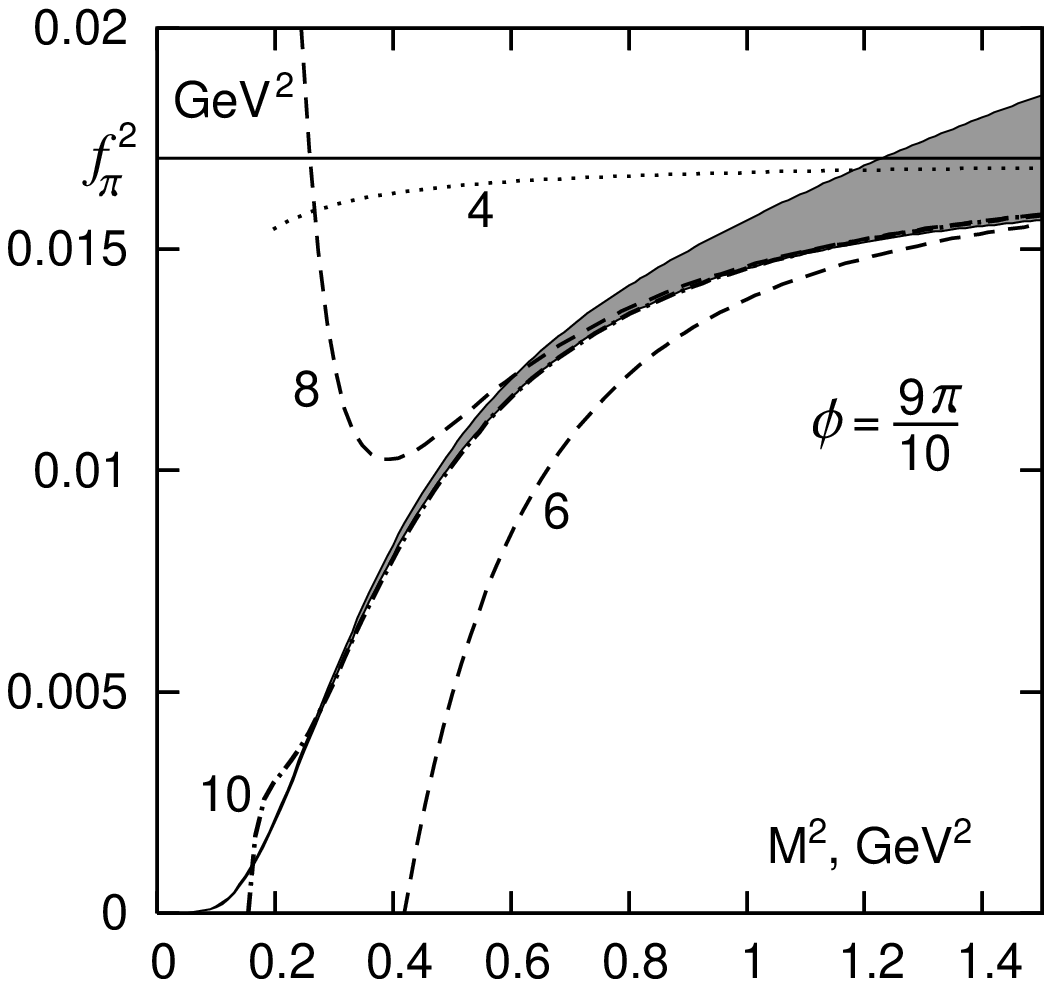, width=70mm}
 \caption{Imaginary part of the Borel transformation (\ref{borim})  at $\phi=4\pi/5$ 
 and real part (\ref{borre}) at $\phi=9\pi/10$. The operator $O_{12}^{V-A}$ vanishes in these sum rules.}
 \label{bwo12}
 \end{figure}

The validity of our assumptions is demonstrated in the Figure \ref{bs_12}. If the operator 
$O_{10}^{V-A}$ is taken into account, a good agreement of theoretical and experimental values is 
observed for $M^2>0.4\,{\rm GeV}^2$. Below this value the contribution of the operator $O_{12}$ 
could be large. Even better agreement can be found at the angles, where the operator $O_{12}$
disappears, see the plots in the Figure \ref{bwo12}. Here the fit can be extended down to
$M^2=0.3\,{\rm GeV}^2$. One may also obtain the condensates by fitting the real part of the Borel
transformation (\ref{borre}). Here the central values of the condensates turns out to be close to
(\ref{combfit}), but the errors are higher due to the presence of additional parameter $f_\pi^2$.
Combined fit of eq (\ref{borre}) at different angles $\phi$ gives $f_\pi=131\pm 4\,{\rm MeV}$. As pointed in 
\cite{IZ}, $f_\pi$ itself has an ambiguity of order $m_\pi^2/m_\rho^2 \sim 3\%$, the accuracy of
the chiral lagrangian parameters. Notice the sign alternation in (\ref{combfit}), in agreement with 
the minimal hadronic ansatz for $\Pi_V-\Pi_A$ correlator, constructed in \cite{PPR} in the 
large $N_c$ limit.

Finally, we write down the values of the quark condensate and the parameter $m_0^2$,
obtained from the operators (\ref{combfit}):
\be
\alpha_s \langle {\bar q} q \rangle^2 \left( 1 + c_6 {\alpha_s\over \pi}\right) \,=\,
 (262\pm 9 \, {\rm MeV} )^6
\label{qcnum}
\ee
\be
 m_0^2\, =\,-O_8^{V-A}/O_6^{V-A} \,=\,1.1\pm 0.2 \, {\rm GeV}^2
\ee
The errors in the r.h.s.~are purely experimental: they do not include possible contribution
of the operator $O_{12}$ and higher as well as unknown QCD corrections to the condensates.
The factor $c_6$ is scheme dependent and left arbitrary in (\ref{qcnum}). The accuracy 
of $m_0^2$ is better than accuracy of $O_8^{V-A}$ because of high covariance of $O_6$ and $O_8$.
Notice a very good agreement of the $D=10$ condensate (\ref{combfit}) obtained from the 
sum rules,  with the one estimated in previous section (\ref{o10num}) within 
the framework of the factorization hypothesis.


\section{Conclusion}

We have performed the analysis of the $V-A$ spectral functions, obtained from hadronic
$\tau$-decay channels, with the help of the Borel sum rules. The values of the 
condensates of dimension $D=6,8,10$ were found (\ref{combfit}) by fitting the theoretical curves 
of the Borel transform to the experimental ones within its errorbands. The major contribution 
to these condensates comes from the 4-quark operators. Its contribution to the
current correlators was calculated and their size was estimated 
by means of the factorizations hypothesis. The estimated value of the $D=10$ condensate
(\ref{o10num}) is found to be in good agreement with the fit result (\ref{combfit}), which 
demonstrates the validity of OPE approach in Quantum Chromodynamics. 

Our results are based on several assumptions, in particular, the factorization (vacuum
insertion) hypothesis. There is a statement in literature \cite{LNT}, that factorization hypothesis
underestimates the quartic condensates by a factor $\sim 3$. This conclusion is based
on the comparison of the quark condensate obtained from the $D=6$ operator in $\rho$-meson
(vector) sum rules with the one calculated from the low-energy GMOR theorem. 
(Our result (\ref{qcnum}) is also larger than GMOR condensate for reasonable theoretical parameters.)
However this comparison has many other sources of error, such as scale-scheme ambiguity, 
high order QCD corrections, light quark masses, corrections from the chiral lagrangian etc.
The accuracy of the factorization hypothesis can be of the same order as the ambiguity of 
the factorization of the D=8 operators the level of $O(N_c^{-2})$ terms, as demonstrated in \cite{IZ}.
More careful statement about validity of the factorization hypothesis could be obtained by 
evaluating the contribution of the meson states to the 4-quark condensates. 

Second objection may concern rather low value of the Borel mass $M_1^2 =0.4\,{\rm GeV}^2$ used
in our fit (\ref{chi2}). Indeed, the typical scale where perturbative results for the current correlators 
are confirmed, is $Q^2 \gtrsim 1\,{\rm GeV}^2$.  But our result for the $D=10$ operator
demonstrates  rather low (power-like) growth of the operators $|O_{2n+2}/O_{2n}|\sim m_0^2$
in the $V-A$ channel. If the operators grow as $|O_{2n}|\sim m_0^{2n}$, then the Borel series 
behaves as $e^{-m_0^2/M^2}$. The contribution of the $n+1$-term in the exponent is small 
for $M^2\gg m_0^2/n$. So for $n=5$ the minimal scale $M_1^2=0.4\,{\rm GeV}^2$ seems reasonable. 
For a faster growth of the operator series this choice could be inappropriate. For instance, if one 
plots the Borel transformation versus $M^2$ with the condensate values obtained in \cite{CGM}, 
the divergence of the operator series will be obvious already at $M^2\approx 0.7\,{\rm GeV}^2$. 
However it should be mentioned, that the $D=10$ condensate obtained there exceeds our 
value (\ref{o10num}) by an order of magnitude. It seems unlikely to explain such discrepancy 
by the inaccuracy of the factorization. All these assumptions can be confirmed or disproved 
only within a nonperturbative approach.

We have neglected the logarithmic terms $\sim \ln{Q^2}/Q^{2n}$ in the OPE series (\ref{disr}).
Such contribution from the $\alpha_s$-correction to the $D=6$ 
condensate (\ref{o6vma}) has a small numerical factor; its discontinuity along the real axis
$Q^2=-s<0$ is too small to compare with the spectral function $v_1(s)-a_1(s)$.
For this reason it would be interesting to calculate the $\alpha_s^2$ correction to the
$D=6$ $V-A$ operator and $\alpha_s$ correction to the operator $O_8^{V-A}$ and 
include them in the sum rule analysis.

\begin{acknowledgement}
Author thanks Sven Menke for providing OPAL datafiles and 
B.L.Ioffe for discussions. This work was supported in part by INTAS grant 2000-587 and 
RFBR grant 03-02-16209. 
\end{acknowledgement}


\section*{Appendix A: 4-quark operators}

\setcounter{equation}{0}
\def\theequation{A\arabic{equation}}

The calculation of the operator contribution to various current correlators can be performed
within the framework of background field method, see for instance \cite{Gr}.
Here we describe the algorithm, conventions and basic formulae,
necessary to calculate the contribution of the 4-quark condensates
to the 2-current correlator, which correspond to the third diagram of the Fig.~\ref{diag_10}. 
We also present here complete form of the 4-quark
operators up to dimension $D=10$.
For definiteness we consider only the vector current correlator; the
condensate contribution to the axial current correlator is trivially obtained by the
substitution $d\to \gamma^5 d$. The contribution of the 4-quark condensates can be written as:
\bea
&& \Pi_{\mu\nu}^V(q) =  -{ig^2\over 4} \int \! dx \, dy \, dz\,  e^{iqx} \biggl< D_{\alpha\beta}^{ab} (y,z)
 \label{4qc2cc} \\
&& \times \biggl[
 \bar{u}(x)\gamma_\mu S(x,y) \gamma_\alpha \lambda^a d(y) \, +
 \bar{u}(y)\gamma_\alpha\lambda^a S(y,x)\gamma_\mu d(x) \biggr]   \nonumber \\
&& \times \biggl[
 \bar{d}(0)\gamma_\nu S(0,z) \gamma_\beta \lambda^b u(z) \, +
 \bar{d}(z)\gamma_\beta\lambda^b S(z,0)\gamma_\nu u(0) \biggr] \biggr> \nonumber 
\eea
Here $S(x,y)=\left< Tq(x){\bar q}(y)\right>$ is the quark Green function
and $D^{ab}_{\mu\nu}(x,y)=\left<Ta^a_\mu(x)a^b_\nu(y)\right>$ is the gluon
Green function in background gluon field $A_\mu\to A_\mu+a_\mu$. They obey the equations:
\be
\label{qgfem}
 i\hat{D}_x \, S(x,y)\,=\,i\delta^4(x-y)
\ee
\be
\label{ggfem}
\left[ D_x^2 g_{\mu\alpha} + 2G_{\mu\alpha}(x) \right]\!{}^{ac}
D^{cb}_{\alpha\nu}(x,y)\,=\,i\delta^{ab} g_{\mu\nu}\delta^4(x-y)
\ee
where $g_{\mu\nu}=(+,-,-,-)$ is Minkowski metric.
The quarks are massless, the gluon Green function is taken in the Feynman gauge. The
covariant derivative and the gluon field-strength tensor
 in the fundamental representation (\ref{qgfem}) are defined as follows:
\bea
&& D_\mu=\partial_\mu - iA_\mu   , \quad
 A_\mu={g\over 2}\lambda^a A_\mu^a   ,  \nonumber \\
&& G_{\mu\nu}=i[D_\mu , D_\nu]={g\over 2}\lambda^a G^a_{\mu\nu}
\eea
where $\lambda^a$ are Gell-Mann matrices ${\rm tr}(\lambda^a\lambda^b)=2\delta^{ab}$,
$[\lambda^a,\lambda^b]$ $=if^{abc}\lambda^c$. We shall also use additional compact
notations for these objects in adjoint representation (\ref{ggfem}):
\be
D_\mu^{ab}=\partial_\mu \delta^{ab} + A_\mu^{ab} , \quad
A_\mu^{ab}={g\over 2}f^{acb}A_\mu^c   , \quad
G_{\mu\nu}^{ab}={g\over 2}f^{acb}G_{\mu\nu}^c
\label{adjr}
\ee

It is convenient to perform partial Fourier transformation of the Green functions:
\bea
S(x,y) & = & \int \!{d^4 q\over (2\pi)^4} \, e^{-iq(x-y)} {\tilde S}(q,y)   , \nonumber \\
D^{ab}_{\mu\nu} (x,y) & = & \int\! {d^4 q\over (2\pi)^4} \, e^{-iq(x-y)} {\tilde D}_{\mu\nu}^{ab} (q,y)
\eea
Then one can write down the solution of equations (\ref{qgfem}), (\ref{ggfem}) as series in
powers of background field $A$:
\bea
\label{tilssol}
 \tilde{S}(q,y) & = & S_0(q)\sum_{n=0}^\infty \left[ i\hat{A}(\hat{x})S_0(q) \right]^n   , \\
{\tilde D}_{\mu\nu}^{ab}(q,y) & = & \Bigl\{ D_0(q)\sum_{n=0}^\infty \left[ iR(q,\hat{x})D_0(q) \right]^n
  \Bigr\}  {}^{ab}_{\mu\nu}, \label{tildsol}
\eea
where
$$
 S_0(q) \,=\, {i\hat{q}\over q^2} , \qquad
 D_0(q)=-{i\over q^2}
$$
are free propagators, ${\hat x}=y-i\overrightarrow{\partial}$, the derivative
$\overrightarrow{\partial}= \partial/\partial q$
acts on everything from the right as $[\overrightarrow{\partial}_\mu, q_\nu ]=g_{\mu\nu}$;
$R$ is the following matrix operator:
\bea
 R_{\mu\nu}^{ab}(q,\hat{x}) & = & \left[-iq_\alpha A_\alpha^{ab}(\hat{x}) -i A_\alpha^{ab}(\hat{x})q_\alpha 
 \right. \nonumber \\
 & & \left. + A_\alpha^{ac}(\hat{x})A_\alpha^{cb}(\hat{x})\right] g_{\mu\nu} + 2G_{\mu\nu}^{ab}(\hat{x})
\eea
The equations (\ref{tilssol}), (\ref{tildsol}) can be evaluated in a
gauge covariant way in the fixed point gauge $x^\mu A_\mu (x)=0$, where
\bea
A_\mu(x)&=&-\int_0^1 d\alpha \, \alpha x^\nu G_{\mu\nu}(\alpha x)  \label{afp} \\
 &=& -x^\nu\sum_{n=0}^\infty {x^{\alpha_1} \ldots x^{\alpha_n}\over (n+2) n!}\,
D_{\alpha_1}\ldots D_{\alpha_n} G_{\mu\nu}(0)  \nonumber \\
q(x)& = &\sum_{n=0}^\infty {x^{\alpha_1} \ldots x^{\alpha_n}\over n!}\, D_{\alpha_1}\ldots D_{\alpha_n} q(0)
\eea
In order to compute the propagators $\tilde{S}$, $\tilde{D}$ for any fixed order $n$, one
has to substitute (\ref{afp}) into (\ref{tilssol}), (\ref{tildsol}),
move all the derivatives $\overrightarrow{\partial}$ to the right  and then leave
only the terms without $\overrightarrow{\partial}$.

The 4-quark condensate contribution (\ref{4qc2cc}) can be written in terms of
the propagators $\tilde{S}$, $\tilde{D}$ as follows:
\bea
\Pi_{\mu\nu}^V(q) & = &  -{ig^2\over 4}\biggl[ 
\bar{u}(-i\overrightarrow{\partial})\gamma_\mu \tilde{S}(q,-i\overrightarrow{\partial})
\gamma_\alpha \lambda^a d(-i\overrightarrow{\partial}) \, \nonumber \\
 & \times & \tilde{D}^{ab}_{\alpha\beta}(q,-i\overrightarrow{\partial})\,X^b_{\nu\beta}(q) 
 + X^b_{\nu\beta}(q)\, \tilde{D}^{ba}_{\beta\alpha}(-q,-i\overleftarrow{\partial}) \nonumber \\
 & \times & \bar{u}(-i\overleftarrow{\partial})\gamma_\alpha\lambda^a \tilde{S}(-q,-i\overleftarrow{\partial})
\gamma_\mu d(-i\overleftarrow{\partial}) \biggr] ,
\eea
where
\bea
\label{xbb}
X_{\nu\beta}^b(q) & =  & \bar{d}(-i\overrightarrow{\partial})\gamma_\beta\lambda^b
\tilde{S}(q,0)\gamma_\nu u(0) \nonumber \\
 & + & \bar{d}(0)\gamma_\nu \tilde{S}(-q,-i\overleftarrow{\partial})
 \gamma_\beta \lambda^b u(-i\overleftarrow{\partial}) \; .
\eea
In the functions $\tilde{S}(q,y)$ and $\tilde{D}(q,y)$ the derivatives $\overrightarrow{\partial}$,
$\overleftarrow{\partial}$ over momentum $q$
always stand on the right from any function of $q$: $\ldots q \ldots \partial$.
The derivatives inside (\ref{xbb}) do not act on anything outside $X^b_{\nu\beta}$.
After these derivatives are evaluated, we compute the transverse part
$\Pi^{(1)}=-\Pi_{\mu\mu}/(3 q^2)$ defined according to (\ref{2ccdef}).
(We also checked, that longitudinal part vanishes $\Pi^{(0)}=0$.)
And finally, to separate out the Lorentz invariant condensates, we average
$\Pi^{(1)}$ over directions of vector $q_\mu$ according to:
\bea
 && \ov{q_{\mu_1} \ldots q_{\mu_{2n}}} =  2{(2n-1)!!\over (2n+2)!!} (q^2)^n
 g_{(\mu_1\mu_2} \ldots g_{\mu_{2n-1}\mu_{2n})}   , \nonumber \\
 & &  \ov{q_{\mu_1} \ldots q_{\mu_{2n+1}}}=0 ,
\eea
where $(\mu_1 \ldots \mu_n)$ denotes usual index symmetrization with weight $1/n!$.
All these calculations were performed on computer.

The most time-consuming part of the calculation is to reduce large number of terms in
the final result to minimal number of independent structures. For this purpose
we employ the quark equation of motion ${\hat D}u={\hat D}d=0$ and
the "integration by part" identity $\langle A(D_\mu B) \rangle = -\langle (D_\mu A) B \rangle$
(the vacuum average of the total derivative is zero
$\langle \partial_x O(x) \rangle= \partial_x \langle O(x) \rangle=\partial_x \langle O(0) \rangle=0$
 for any gauge invariant operator $O(x)$). It allows to bring the operators to
obviously hermitean (real) form, which provides an additional verification of the result.

In order to write down the 4-quark condensates in a compact form, we introduce
here the following bilinear quark structures of increasing dimension $D$:
\be
\ba{ll}
3D: \quad & A_\alpha =  {\bar d}\lambda\gamma^5\gamma_\alpha u\\
4D: & B^{(1)}_{\alpha\beta} = i({\bar d}\lambda\gamma_\alpha u_\beta
 - {\bar d}_\beta\lambda\gamma_\alpha u) \\
 & B^{(2)}_{\alpha\beta} = {\bar d}\lambda\gamma^5\gamma_\alpha u_\beta
 + {\bar d}_\beta\lambda\gamma^5\gamma_\alpha u\\
5D: & C^{(1)}_{\alpha\beta\gamma}=i({\bar d}\lambda\gamma_\alpha u_{\beta\gamma}
 - {\bar d}_{\beta\gamma} \lambda\gamma_\alpha u) \\
 & C^{(2)}_{\alpha\beta\gamma} ={\bar d}\lambda\gamma^5\gamma_\alpha u_{\beta\gamma}
 + {\bar d}_{\beta\gamma}\lambda\gamma^5\gamma_\alpha u\\
 & C^{(3)}_{\alpha\beta\gamma} = {\bar d}_\alpha\lambda\gamma^5\gamma_\beta u_\gamma
 + {\bar d}_\gamma\lambda\gamma^5\gamma_\beta u_\alpha\\
 & C^{(4)}_{\alpha\beta\gamma} ={\bar d}\{ \lambda, G_{\alpha\beta}\}\gamma_\gamma u \\
6D: & E^{(1)}_{\alpha\beta\gamma\delta}= {\bar d}\{\lambda,G_{\alpha\beta}\}\gamma_\gamma u_\delta
 + {\bar d}_\delta \{\lambda, G_{\alpha\beta}\}\gamma_\gamma u \\
& E^{(2)}_{\alpha\beta\gamma\delta}= {\bar d}\{\lambda, \tilde{G}_{\alpha\beta}\}\gamma_\gamma u_\delta
 + {\bar d}_\delta\{\lambda, \tilde{G}_{\alpha\beta}\}\gamma_\gamma u \\
& E^{(3)}_{\alpha\beta\gamma\delta}= i( {\bar d}\{\lambda, G_{\alpha\beta}\}\gamma^5\gamma_\gamma
 u_\delta  -  {\bar d}_\delta \{\lambda, G_{\alpha\beta}\} \gamma^5 \gamma_\gamma u ) \\
& E^{(4)}_{\alpha\beta\gamma\delta}= i( {\bar d}\{\lambda,\tilde{G}_{\alpha\beta}\}\gamma^5
 \gamma_\gamma u_\delta - {\bar d}_\delta \{\lambda, \tilde{G}_{\alpha\beta}\}\gamma^5 \gamma_\gamma u) \\
& E^{(5)}_{\alpha\beta\gamma\delta}= {\bar d}\{\lambda, G_{\beta\gamma;\,\alpha}\} \gamma_\delta u  \\
& E^{(6)}_{\alpha\beta\gamma\delta}= {\bar d}\{\lambda, \tilde{G}_{\beta\gamma;\,\alpha}\}\gamma_\delta u \\
7D: & F^{(1)}_{\alpha\beta\gamma}=i( {\bar d}\{\lambda, J_\alpha\} \gamma^5\gamma_\beta u_\gamma -
    {\bar d}_\gamma \{\lambda, J_\alpha\} \gamma^5 \gamma_\beta u ) \\
& F^{(2)}_{\alpha\beta\gamma} ={\bar d}\{\lambda ,\{ G_{\alpha\delta} , G_{\delta\beta} \} \}
   \gamma^5\gamma_\gamma u \\
& F^{(3)}_{\alpha\beta\gamma} =i{\bar d}\{\lambda , [G_{\alpha\delta} , \tilde{G}_{\delta\beta} ] \}
   \gamma_\gamma u
\ea
\label{bistr}
\ee
where $u_\alpha=D_\alpha u$, $u_{\alpha\beta}=D_{(\alpha}D_{\beta)}u\equiv 
 {1\over 2}(D_\alpha D_\beta+D_\beta D_\alpha)u$,
 $G_{\beta\gamma;\,\alpha}=D_\alpha G_{\beta\gamma}$, $J_\alpha=D_\beta G_{\alpha\beta}$,
  $[A,B]=AB-BA$, $\{A,B\}=AB+BA$. The dual tensor is defined as
 $\tilde{G}_{\alpha\beta}={1\over 2}\ve_{\alpha\beta\mu\nu} G_{\mu\nu}$,
 $\ve^{0123}=1$ and $\gamma^5=i\gamma^0\gamma^1\gamma^2\gamma^3$.
  The values $G$, $\tilde{G}$, $J$ in (\ref{bistr}) are in
 fundamental representation. All bilinear structures belong to adjoint representation of the gauge group,
 the gauge index  of the Gell-Mann matrices $\lambda$ is omitted.
We denote conjugated structures by overlined letters, which are simply obtained
by the replacement $u \rightleftarrows d$, for instance
${\bar A}_\alpha \equiv A_\alpha^\dagger = {\bar u}\lambda\gamma^5\gamma_\alpha d$.

The 4-quark condensates of dimension $D=6,8,10$ are:
\bea
 && O_6^V  =  -2\pi\alpha_s \left< {\bar A}_\alpha A_\alpha \right>  \label{o6v} \\
 && O_8^V  =  {2\pi\alpha_s\over 9}\left<
 -4{\bar B}^{(1)}_{\alpha\beta}B^{(1)}_{\alpha\beta}
 - {\bar B}^{(2)}_{\alpha\beta}B^{(2)}_{\alpha\beta} 
 \right. \nonumber \\ & & \qquad \quad \left. 
 -4{\bar C}^{(3)}_{\beta\alpha\beta}A_\alpha
 -4{\bar A}_\alpha C^{(3)}_{\beta\alpha\beta}
 + 12 {\bar A}_\alpha G_{\alpha\beta} A_\beta \right>
 \label{o8v} \\
 && O_{10}^V  =  {\pi\alpha_s\over 9} \Biggl<
  25 {\bar C}^{(1)}_{\alpha\beta\gamma} C^{(1)}_{\alpha\beta\gamma}
 - 5 {\bar C}^{(2)}_{\alpha\beta\gamma} C^{(2)}_{\alpha\beta\gamma}
 - 10 {\bar C}^{(3)}_{\alpha\beta\alpha} C^{(3)}_{\gamma\beta\gamma}
\nonumber \\  & & 
 - 19 {\bar C}^{(4)}_{\alpha\beta\beta} C^{(4)}_{\alpha\gamma\gamma}
 - {15\over 4} {\bar C}^{(4)}_{\alpha\beta\gamma} C^{(4)}_{\alpha\beta\gamma}
 - 8 {\bar C}^{(4)}_{\alpha\beta\gamma} C^{(4)}_{\beta\gamma\alpha}
\nonumber \\  & &  
- 2 {\bar B}^{(1)}_{\alpha\beta} G_{\beta\gamma} B^{(1)}_{\alpha\gamma}
 - 66 {\bar B}^{(2)}_{\alpha\beta} G_{\alpha\gamma} B^{(2)}_{\gamma\beta}
  + {\bar A}_\alpha \Bigl(  8 J_{[\alpha;\,\beta]}
\nonumber \\ & & 
 - 3 G_{\alpha\gamma} G_{\gamma\beta}
 + 19 G_{\beta\gamma} G_{\gamma\alpha} \Bigr)  A_\beta
 + {33\over 4} {\bar A}_\alpha G_{\beta\gamma} G_{\beta\gamma} A_\alpha
 \nonumber \\  & &
 + {\bar B}^{(1)}_{\alpha\beta} \left(
    E^{(1)}_{\beta\gamma\alpha\gamma}
   + {5\over 2} E^{(4)}_{\alpha\gamma\gamma\beta}
   - {7\over 2} E^{(5)}_{\gamma\beta\gamma\alpha}
   - 28 {\tilde G}_{\beta\gamma} B^{(2)}_{\alpha\gamma} 
 \right. \nonumber \\  & & \left.   
+ {21\over 2} {\tilde G}_{\beta\gamma;\,\alpha} A_\gamma \right) 
 + \left(
   {\bar E}^{(1)}_{\beta\gamma\alpha\gamma}
  + {5\over 2} {\bar E}^{(4)}_{\alpha\gamma\gamma\beta}
  - {7\over 2} {\bar E}^{(5)}_{\gamma\beta\gamma\alpha} 
 \right. \nonumber \\  & &  \left.
  + 28 {\bar B}^{(2)}_{\alpha\gamma} {\tilde G}_{\beta\gamma}
  - {21\over 2} {\bar A}_\gamma {\tilde G}_{\beta\gamma;\,\alpha} \right)  B^{(1)}_{\alpha\beta} 
 + {\bar B}^{(2)}_{\alpha\beta} \left(
  - {11\over 2} E^{(2)}_{\alpha\gamma\gamma\beta} 
\right. \nonumber \\  & & \left. 
  + {15\over 4} E^{(2)}_{\beta\gamma\alpha\gamma}
  + {5\over 2} E^{(3)}_{\alpha\gamma\beta\gamma}
  + 5 E^{(3)}_{\beta\gamma\alpha\gamma}
  + {1\over 2} E^{(6)}_{\alpha\beta\gamma\gamma}
  + {3\over 2} E^{(6)}_{\beta\alpha\gamma\gamma}
 \right. \nonumber \\  & & \left.
  - 4 E^{(6)}_{\gamma\alpha\beta\gamma}
   - {17\over 2} G_{\alpha\beta;\,\gamma} A_{\gamma}
  - 11 J_\beta A_\alpha \right)
  + \left( - {11\over 2} {\bar E}^{(2)}_{\alpha\gamma\gamma\beta}
  \right. \nonumber \\  & & \left.
  + {15\over 4} {\bar E}^{(2)}_{\beta\gamma\alpha\gamma}
  + {5\over 2} {\bar E}^{(3)}_{\alpha\gamma\beta\gamma}
  + 5 {\bar E}^{(3)}_{\beta\gamma\alpha\gamma}
  + {1\over 2} {\bar E}^{(6)}_{\alpha\beta\gamma\gamma}
  + {3\over 2} {\bar E}^{(6)}_{\beta\alpha\gamma\gamma}
  \right. \nonumber \\  & & \left.
  - 4 {\bar E}^{(6)}_{\gamma\alpha\beta\gamma}
  + {17\over 2} {\bar A}_\gamma G_{\alpha\beta;\,\gamma}
   + 11 {\bar A}_\alpha J_\beta \right)  B^{(2)}_{\alpha\beta}
 \nonumber \\  & & 
+ {\bar A}_\alpha \left(
  - 3 F^{(1)}_{\beta\alpha\beta}
  + F^{(1)}_{\beta\beta\alpha}
  + 2 F^{(2)}_{\alpha\beta\beta}
  - {1\over 2} F^{(2)}_{\beta\beta\alpha}
  +  F^{(3)}_{(\alpha\beta)\beta}
  \right. \nonumber \\  & & \left.
- {45\over 2} {\tilde G}_{\alpha\beta} C^{(4)}_{\beta\gamma\gamma}
  + {35\over 2} {\tilde G}_{\beta\gamma} C^{(4)}_{\alpha\beta\gamma}
  + {75\over 8} {\tilde G}_{\beta\gamma} C^{(4)}_{\beta\gamma\alpha} \right) 
 \nonumber \\ & &
 + \left(
 -3 {\bar F}^{(1)}_{\beta\alpha\beta}
 + {\bar F}^{(1)}_{\beta\beta\alpha}
  + 2 {\bar F}^{(2)}_{\alpha\beta\beta}
  - {1\over 2} {\bar F}^{(2)}_{\beta\beta\alpha}
  +  {\bar F}^{(3)}_{(\alpha\beta)\beta}
 \right. \nonumber \\  & & \left.
  + {45\over 2} {\bar C}^{(4)}_{\beta\gamma\gamma} {\tilde G}_{\alpha\beta}
  - {35\over 2} {\bar C}^{(4)}_{\alpha\beta\gamma} {\tilde G}_{\beta\gamma}
  - {75\over 8} {\bar C}^{(4)}_{\beta\gamma\alpha} {\tilde G}_{\beta\gamma} \right) \!  A_\alpha \!
 \Biggr>\nonumber \\
 & & 
 \label{o10v}
\eea
In (\ref{o8v}), (\ref{o10v}) the field strengths $G$, $\tilde{G}$, $J$ are in adjoint representation
$G_{\alpha\beta}^{ab}={g\over 2} f^{acb} G^c_{\alpha\beta}$ etc;
gauge indices  are omitted, say ${\bar A}_\alpha G_{\beta\gamma} G_{\beta\gamma} A_\alpha$ denotes
${\bar A}_\alpha^a G_{\beta\gamma}^{ab} G_{\beta\gamma}^{bc} A_\alpha^c$.
The operator $O_8^V$ (\ref{o8v}) can be easily brought to the form, obtained in \cite{IZ} and \cite{GP}.


\section*{Appendix B: Factorization of 4-quark condensates}

\setcounter{equation}{0}
\def\theequation{B\arabic{equation}}

At first let us remind, how the factorization (vacuum insertion) works for the $D=6$ operators.
It is illustrated by the following equation:
\be
\langle ({\bar u}\lambda \Gamma_1 d)({\bar d} \lambda \Gamma_2 u)\rangle \,=\,
 -2\, C_N \, {\rm tr}\left[ \langle u \otimes {\bar u} \rangle \Gamma_1
  \langle d \otimes {\bar d} \rangle \Gamma_2 \right]
\label{4qfac}
\ee
where $\Gamma_i$ are some Dirac matrices, $C_N=1-1/N_c^2$, $N_c$ is the color number,
kept arbitrary here. In (\ref{4qfac}) the notation $\langle q \otimes {\bar q}\rangle$ denotes $4\times 4$
matrix in spinor space, the color indices are contracted. It is proportional to the quark condensate:
\be
\left< q \otimes {\bar q}\right> \,=\, -\,{1\over 4}\,\langle {\bar q}q\rangle
\ee
The result of the factorization is well known:
\be
O^V_6\,=\,-4\pi\alpha_s C_N \langle {\bar u}u \rangle  \langle {\bar d}d \rangle
\ee
(In the vector sum rules one also accounts additional operator 
$\langle {\bar q} \gamma_\alpha J_\alpha q \rangle$, which takes the 4-quark form
when the gluon equation of motion is applied. Such operator comes from the 2-quark diagram, 
so it cancels in the $V-A$ correlators.)

The factorization procedure becomes ambiguous at the level of $D=8$ 4-quark condensates.
As shown in \cite{IZ}, different ways of factorization give different terms $\sim 1/N_c^2$. 
For definiteness, here we follow the following factorization scheme.
At first we replace the field strength by the derivatives as $G_{\mu\nu}=i[D_\mu , D_\nu]$ for
fundamental representation and $G_{\mu\nu}^{ab}=[D_\mu, D_\nu]^{ab}$ for adjoint one.
Then we apply the equation (\ref{4qfac}), where the quark wave functions $u,{\bar u}$ and $d,{\bar d}$
may carry some derivatives. Finally, the quark matrices $\langle \ldots\rangle$ with derivatives are 
expressed in terms of the condensates as:
\bea
 && \left<  D_\alpha q \otimes{\bar q}\right> =0 , \nonumber \\
 &&  \left<  D_\alpha D_\beta q \otimes{\bar q}\right> =-{1\over 32} \left( g_{\alpha\beta} +
 {1\over 3}\gamma_{\alpha\beta} \right) i\langle {\bar q} \hat{G} q \rangle  ,
\eea
where $\gamma_{\alpha\beta}=\gamma_{[\alpha}\gamma_{\beta]}={1\over 2}
 (\gamma_\alpha\gamma_\beta - \gamma_\beta\gamma_\alpha) $,
$\hat{G}=\gamma_{\alpha\beta}G_{\alpha\beta}$. The result for the $D=8$ condensate is:
\be
O_8^V\,=\,-2\pi\alpha_s C_N \left[ \, \langle {\bar u}u \rangle\, i \langle {\bar d}\hat{G}d \rangle\, +
 \,\langle {\bar d}d \rangle\, i \langle {\bar u}\hat{G}u \rangle\, \right]
\ee

In the condensate $O_{10}$ one encounters the terms with quarks carrying 4 derivatives. 
We average these terms with the help of the following rule:
\bea
 && \left<  D_\alpha D_\beta D_\gamma D_\delta q \otimes{\bar q}\right> \,=\,
 - {1\over 24^2}\bigl[  
 g_{\alpha\beta} g_{\gamma\delta} ( 3 X_2 + 6 X_3 - 2 X_4 ) 
\nonumber \\ & & 
 + g_{\alpha\gamma} g_{\beta\delta} ( 6 X_1 + 3 X_2 + 6 X_3 + 4 X_4 ) 
 + g_{\alpha\delta} g_{\beta\gamma} ( 12 X_1 
\nonumber \\  & & 
 + 3 X_2 + 6 X_3 + 4 X_4 )
 +  ( g_{\alpha\beta} \gamma_{\gamma\delta} + g_{\gamma\delta} \gamma_{\alpha\beta}) ( X_1 + 2 X_2 
\nonumber \\  & & 
 + 3 X_3 ) + ( g_{\alpha\gamma} \gamma_{\beta\delta} + g_{\beta\delta} \gamma_{\alpha\gamma})
 ( 2 X_1 + X_2 + 3 X_3 + X_4 )
\nonumber \\  & & 
  + g_{\alpha\delta} \gamma_{\beta\gamma} ( X_1 + 2 X_2 + X_4 )
  + g_{\beta\gamma} \gamma_{\alpha\delta} ( X_1 + X_4 )
\nonumber \\  & & 
 + 3 \gamma_{\alpha\beta\gamma\delta} X_2 \bigr]
\eea
where $\gamma_{\alpha\beta\gamma\delta}=\gamma_{[\alpha}\gamma_\beta
 \gamma_\gamma\gamma_{\delta]}$, $X_i$ are 7-dimensional condensates, defined
in (\ref{d7cond}).

Being applied to the operator (\ref{o10v}), this procedure gives the following result:
\bea
O_{10}^V &  = & \pi\alpha_s C_N \left[ \,
 {25\over 9}\, \langle {\bar u}\hat{G}u \rangle \langle {\bar d}\hat{G}d \rangle\ \, - \right. 
 \nonumber \\  & & 
  - \, 4 \left( 3 X_1^u - X_2^u + X_3^u + {7\over 6}X_4^u  \right) \langle {\bar d}d \rangle 
\nonumber \\  & & \left.
  -\, 4  \left(  3 X_1^d - X_2^d + X_3^d + {7\over 6}X_4^d  \right) \langle {\bar u}u \rangle \, \right]
\eea
where $X_i^q$ are constructed from the quark of flavor $q$. The axial condensates can be obtained 
by simple replacement $d\to \gamma^5 d$. For all factorized 4-quark operators $O_D^A=-O_D^V$.


\begin{thebibliography}{99}
\bibitem{SVZ}
M.A. Shifman, A.I. Vainstein, V.I. Zakharov, Nucl. Phys. B {\bf 147}, 385, 448 (1979)
\bibitem{ALEPH}
ALEPH collaboration: R. Barate et al, Eur. J. Phys. C {\bf 4}, 409 (1998) 
\bibitem{OPAL}
OPAL collaboration: K. Ackerstaff et al, Eur. J. Phys. C {\bf 7}, 571 (1999)
\bibitem{IZ}
B.L. Ioffe, K.N. Zyablyuk, Nucl. Phys. A {\bf 687}, 437 (2001)
\bibitem{BNP}
E. Braaten, S. Narison, A. Pich, Nucl. Phys. B {\bf 373}, 581 (1992)
\bibitem{DGHS}
M. Davier, L. Girlanda, A. Hocker, J. Stern, Phys. Rev. D {\bf 58}, 096014 (1998)
\bibitem{PPR} 
S. Peris, B. Phily, E. de Rafael, Phys. Rev. Lett. {\bf 86}, 14 (2001)
\bibitem{BGP}
J. Bijnens, E. Gamiz, J.Prades, JHEP {\bf 0110}, 009 (2001) 
\bibitem{CGM}
V. Cirigliano, E. Golowich, K. Maltman, Phys. Rev. D {\bf 68}, 054013 (2003) 
\bibitem{DomS}
C.A. Dominguez, A. Schilcher, Phys. Lett B {\bf 581}, 193 (2004)
\bibitem{CSSS}
S. Ciulli, C. Sebu, K. Schilcher, H. Spiesberger, hep-ph/0312212
\bibitem{RL}
J. Rojo, J.I. Latorre, JHEP {\bf 0401}, 055 (2004)
\bibitem{Gen}
S.C. Generalis, J. Phys. G {\bf 15}, L225 (1989)
\bibitem{CGS}
K.G. Chetyrkin, S.G. Gorishny, V.P. Spiridonov, Phys. Lett. B {\bf 160}, 149 (1985)
\bibitem{GMOR}
M. Gell-Mann, R.J. Oakes, B. Renner, Phys. Rev. {\bf 175}, 2195 (1968)
\bibitem{LSC}
L.V. Lanin, V.P. Spiridonov, K.G. Chetyrkin, Yad. Fiz. {\bf 44}, 1372 (1986)
\bibitem{AC}
L.-E. Adam, K.G. Chetyrkin, Phys. lett. B {\bf 329}, 129 (1994)
\bibitem{DS}
M.S. Dubovikov, A.V. Smilga, ITEP-82-42
\bibitem{GP}
A. Grozin, Y. Pinelis, Phys. Lett. B {\bf 166}, 429 (1986)
\bibitem{BI}
V.M. Belyaev, B.L. Ioffe, Sov. Phys. JETP {\bf 56}, 493 (1982)
\bibitem{DJN}
H.G. Dosch, M. Jamin, S. Narison, Phys. Lett. B {\bf 220}, 251 (1989)
\bibitem{Nar}
S. Narison, Phys. Lett. B {\bf 210}, 238 (1988)
\bibitem{ChH}
T.-W. Chiu, T.-H. Hsieh, Nucl. Phys. B {\bf 673}, 217 (2003)
\bibitem{DiGS}
A.Di Giacomo, Yu.A. Simonov, hep-ph/0404044
\bibitem{IZ2}
B.L. Ioffe, K.N. Zyablyuk, Eur. Phys. J. C {\bf 27}, 229 (2003)
\bibitem{LNT}
G. Launer, S. Narison, R. Tarrach, Z. Phys. C {\bf 26}, 433 (1984)
\bibitem{Gr}
A.G. Grozin, Int. J. Mod. Phys. A {\bf 10}, 3497 (1995)
\end{thebibliography}
\end{document}